\documentclass[preprintnumbers,amsmath,amssymb,nofootinbib,superscriptaddress,floatfix,showkeys,twocolumn]{revtex4-1}
\usepackage{color,amsmath,amssymb,graphicx,latexsym,subfigure}
\usepackage{threeparttable,txfonts}
\usepackage{appendix}
\usepackage{ulem}
\usepackage{float}
\usepackage{graphicx}
\usepackage{xspace}

\newcommand{\gev}{\ensuremath{\,\mathrm{GeV}}\xspace}
\newcommand{\tev}{\ensuremath{\,\mathrm{TeV}}\xspace}

\begin{document}

\title{Low energy SUSY confronted with new measurements of W-boson mass and muon $g-2$}

\author{Jin Min Yang}
\email{jmyang@itp.ac.cn}
\affiliation{
 CAS Key Laboratory of Theoretical Physics, Institute of Theoretical Physics, Chinese Academy of Sciences, Beijing 100190, P. R. China \\[0.5em]
}

\affiliation{
 School of Physical Science, University of Chinese Academy of Sciences,  Beijing 100049, P. R. China \\[0.5em]
}
\author{Yang Zhang}
\email{zhangyangphy@zzu.edu.cn}
\affiliation{
 School of Physics, Zhengzhou University, Zhengzhou 450000, P. R. China\\
}
\affiliation{
 CAS Key Laboratory of Theoretical Physics, Institute of Theoretical Physics, Chinese Academy of Sciences, Beijing 100190, P. R. China \\[0.5em]
}

\begin{abstract}
The new CDF II measurement of $W$-boson mass shows a 7$\sigma$ deviation from the Standard Model (SM) prediction, while the recent FNAL measurement of the muon $g-2$ shows a 4.2$\sigma$ deviation (combined with the BNL result) from the SM. Both of them strongly indicate new physics beyond the SM.  
In this work we study the implication of both measurements on low energy supersymmetry. With an extensive exploration of the parameter space of the minimal supersymmetric standard model (MSSM), we find that in the parameter space allowed by current experimental constraints from colliders and dark matter detections, the MSSM can simultaneously explain both measurements 
on the edge of $2\sigma$ level, taking theoretical uncertainties into consideration
. The favored parameter space, characterized by 
a compressed spectrum between bino, wino and stau, with the stop being around 1 TeV, may be covered in the near future LHC searches.  

\end{abstract}

\date{\today}

\maketitle

\section{Introduction}
After the discovery of Higgs boson, the main task of high energy physics is to precisely test the Standard Model (SM) and probe new physics beyond the SM. For the probe of new physics, the direct search of new particles is obviously the best way, which of course needs rather high energy. So far only the Large Hadron Collider (LHC) has such an ability for directly probing TeV scale new physics and the high luminosity LHC (HL-LHC) will continue the task of direct searches of new particles. On the other hand, new physics will contribute through loops to electroweak observables like the W-boson mass and the muon anomalous magnetic moment ($g-2$). Hence the precision measurements of electroweak observables and the muon $g-2$ can reveal the quantum effects of new physics and allow for an indirect probe. 

Although direct searches of new physics at the LHC have merely given disappointing results to date,  two exciting harbingers of new physics have appeared, one is the measurement of  muon $g-2$ and the other is the measurement of W-boson mass:
\begin{itemize}
\item The muon $g-2$ from the recent FNAL measurement~\cite{Muong-2:2021ojo}, combined with the BNL result, has a value 
which is $4.2\sigma$ above the SM prediction~\cite{Aoyama:2012wk,Aoyama:2019ryr,Czarnecki:2002nt,Gnendiger:2013pva,Davier:2017zfy,Keshavarzi:2018mgv,Colangelo:2018mtw,Hoferichter:2019mqg,Davier:2019can,Keshavarzi:2019abf,Kurz:2014wya,Melnikov:2003xd,Masjuan:2017tvw,Colangelo:2017fiz,Hoferichter:2018kwz,Gerardin:2019vio,Bijnens:2019ghy,Colangelo:2019uex,Blum:2019ugy,Colangelo:2014qya,Aoyama:2020ynm}
\begin{equation} \label{eq:delta-amu}
    \Delta a_{\mu}^{\rm{Exp-SM}} =a_\mu^{\rm Exp}-a_\mu^{\rm SM}= (2.51\pm 0.59) \times 10^{-9}.
\end{equation}
\item The W-boson mass measured by CDF II is about $7\sigma$ above the SM prediction \cite{CDF:2022hxs}
\begin{eqnarray} 
M_W^{\rm CDF} &=& 80433.5\pm 9.4 {\rm ~MeV}  \\
M_W^{\rm SM} &=& 80357\pm 4 \pm 4 {\rm ~MeV} 
\end{eqnarray}
Remarkably, the precision of such a new measurement exceeds all previous measurements combined.
\end{itemize} 
While there may be some debates about the robustness of these measurements (especially the CDF II W-mass measurement), they may indeed indicate existence of new physics around TeV scale. Among various new physics theories in the literature, the most popular candidate for TeV-scale new physics has been the low energy supersymmetry (SUSY)  (for recent reviews see, e.g., \cite{Athron:2022uzz,Wang:2022rfd,Baer:2020kwz}).  There are numerous motivations for low energy SUSY, in which the most appealing ones are the solution or alleviation of the naturalness problem, the natural explanation of dark matter (DM) and the realization of gauge coupling unification. All these virtues may not be just a coincidence and instead they suggest that low energy SUSY may be the true story of nature. Actually,  the 125 GeV Higgs boson discovered by the LHC is within the mass range predicted by low energy SUSY, which implies heavy stops above TeV (heavy stops are needed to push up the SM-like Higgs boson to 125 GeV). So the un-observation of any sparticles at the LHC, which has pushed up colored sparticles above TeV, is just consistent with the discovery of the 125 GeV Higgs boson in the framework of SUSY.  

Confronted with the two deviations, i.e. the W-boson mass and muon $g-2$, what is the situation of low energy SUSY ?  In this work we restrain our analysis to the minimal supersymmetric standard model (MSSM) which has the most economic particle content. We note that recently there are many studies on the MSSM contributions to muon $g-2$ and the MSSM parameter space allowed by other constraints (including DM) is shown to be able to explain the muon $g-2$ at $2\sigma$ level (see, e.g.,  \cite{Athron:2021iuf, Wang:2021bcx, Li:2021pnt,Chakraborti:2021bmv,Endo:2021zal,Abdughani:2019wai,Han:2021ify,Li:2021xmw,Li:2021koa,Abdughani:2021pdc,Cao:2021tuh,Chakraborti:2021dli,Chakraborti:2021kkr,Chakraborti:2021mbr,Bagnaschi:2022qhb,Cao:2022htd}). Interestingly, the light electroweakinos and sleptons required for the muon $g-2$ explanation in the MSSM are partially accessible at the HL-LHC \cite{Abdughani:2019wai}.   
Confronted with the latest CDF II measurement of the W-boson mass, we should examine the MSSM parameter space allowed by current collider and DM experiments to figure out whether this model can accommodate this new measured $M_W$ value. This is the aim of this work. 

This work is organized as follows. In Section \ref{sec:mw} we briefly revisit the W-boson mass prediction in the MSSM. In Section \ref{sec:strategy} we describe our strategy for numerical analysis. The results are displayed in Section \ref{sec:result}.
Finally, we give a summary in Section \ref{sec:conclusion}.  

\section{W-boson mass in MSSM}\label{sec:mw}

In the SM and its extensions  the W-boson mass can be evaluated from~\cite{Ellis:2007fu}
\begin{equation}
    M_W^2 (1-\frac{M_W^2}{M_Z^2}) = \frac{\pi\alpha}{\sqrt{2}G_{F}} (1+\Delta r),
\end{equation}
where $G_{F}$ is the Fermi constant, $\alpha$ is the fine structure constant, and $\Delta r$ 
represents the sum of all non-QED loop diagrams to the muon-decay amplitude which itself depends on $M_W$ as well. 
This relation arises from comparison between the prediction for muon decay with the Fermi constant.
The one-loop contributions to $\Delta r$ consist of the W-boson self-energy, vertex and box diagrams, and the corresponding counter terms, and can be written conventionally as~\cite{Heinemeyer:2004gx}
\begin{equation}
    \Delta r = \Delta \alpha - \frac{\cos^2\theta_W}{\sin^2\theta_W} \Delta \rho + \cdots,
\end{equation}
where $\Delta \alpha$ is the shift in the fine structure constant arising from the charge renormalization which contains the contributions from light fermions, and $\Delta \rho $  is the loop corrections to the $\rho$ parameter which describes the ratio between neutral and charged weak currents. Approximately, we have 
\begin{equation}
    \delta M_W \simeq \frac{M_W}{2} \frac{\cos^2\theta_W}{\cos^2\theta_W-\sin^2\theta_W} \Delta \rho.
\end{equation}

In the MSSM, the dominant correction at the one-loop level arises from the stop and sbottom contribution 
\begin{align}
    \Delta \rho_0^{\rm SUSY} =  \frac{3G_F}{8\sqrt{2}\pi^2} [
    & - \sin^2 \theta_{\tilde{t}}\cos^2 \theta_{\tilde{t}} F_0 (m^2_{\tilde{t}_1}, m^2_{\tilde{t}_2}) \nonumber \\
    & - \sin^2 \theta_{\tilde{b}}\cos^2 \theta_{\tilde{b}} F_0 (m^2_{\tilde{b}_1}, m^2_{\tilde{b}_2}) \nonumber \\
    & +\cos^2 \theta_{\tilde{t}}\cos^2 \theta_{\tilde{b}} F_0 (m^2_{\tilde{t}_1}, m^2_{\tilde{b}_1}) \nonumber\\
    &  +\cos^2 \theta_{\tilde{t}}\cos^2 \theta_{\tilde{b}} F_0 (m^2_{\tilde{t}_1}, m^2_{\tilde{b}_2}) \nonumber  \\
    & + \sin^2 \theta_{\tilde{t}}\cos^2 \theta_{\tilde{b}} F_0 (m^2_{\tilde{t}_2}, m^2_{\tilde{b}_1}) \nonumber \\
    & + \sin^2 \theta_{\tilde{t}}\cos^2 \theta_{\tilde{b}} F_0 (m^2_{\tilde{t}_2}, m^2_{\tilde{b}_2})
    ]
    \label{eq:stop_contr}
\end{align}
where 
\begin{equation}
    F_0(x,y) = x+y -\frac{2xy}{x-y} \log\frac{x}{y}.
\end{equation}
As $F_0(m^2,m^2) =0$ and $F_0(m^2,0) =m^2$, $\Delta \rho $ is sensitive to the mass splitting between the isospin partners. The contribution of pure slepton loops is similar to squrak loops, while the contributions of the chargino and neutralino sector enter also via vertex and box diagrams.

\section{Strategy of numerical analysis}\label{sec:strategy}

We investigate the prediction for $M_W$ in the MSSM by performing scan of the parameter space, considering relevant experimental constraints. We employ \texttt{SUSY-HIT-1.5}~\cite{Djouadi:2006bz} to generate
the mass spectrum and decay tables of sparticles, and then use the package  \texttt{FeynHiggs-2.18.1}~\cite{Heinemeyer:1998yj,Heinemeyer:2013dia,Bahl:2018qog,Bahl:2016brp,Hahn:2013ria,Frank:2006yh,Degrassi:2002fi,Heinemeyer:1998np} to recalculate the Higgs and sparticles masses, as well as electroweak precision observables including the $W$-bosom mass $M_W^{\rm SUSY}$
\footnote{We discard samples that \texttt{FeynHiggs-2.18.1} gives an warning about questionable $A_b$ from $\overline{\rm DR}$ to $\overline{\rm OS}$ conversion, which may give untruest large $M_W^{\rm SUSY}$.}
, the loop corrections to the $\rho$ parameter  $\Delta \rho$, the effective leptonic weak mixing angle at the $Z$-resonance, $sin^2\hat{\theta}(M_Z)$. The muon $g-2$ is further evaluated by \texttt{GM2Calc}~\cite{Athron:2015rva}. We also employ \textsf{SuperIso v4.1}~\cite{Mahmoudi:2007vz,Mahmoudi:2008tp} to calculate the flavor physics  observables, and \texttt{MicrOMEGAs-5.2.13}~\cite{Belanger:2010pz} to calculate DM relic density, direct and indirect detection cross sections.  The packages \texttt{HiggsBounds-5.10.2}~\cite{Bechtle:2020pkv} and \texttt{HiggsSignals-2.6.2}~\cite{Bechtle:2020uwn} are used to apply constraints at colliders on SM-like and extra Higgs bosons. The SUSY direct searches at LHC are implemented by \texttt{SModelS-2.1.1}~\cite{Kraml:2013mwa,Ambrogi:2017neo,Ambrogi:2018ujg} and \texttt{CheckMATE-2.0.26}~\cite{Drees:2013wra,Dercks:2016npn}.

\begin{figure*}[ht]
 \centering
 \includegraphics[width=0.98\textwidth]{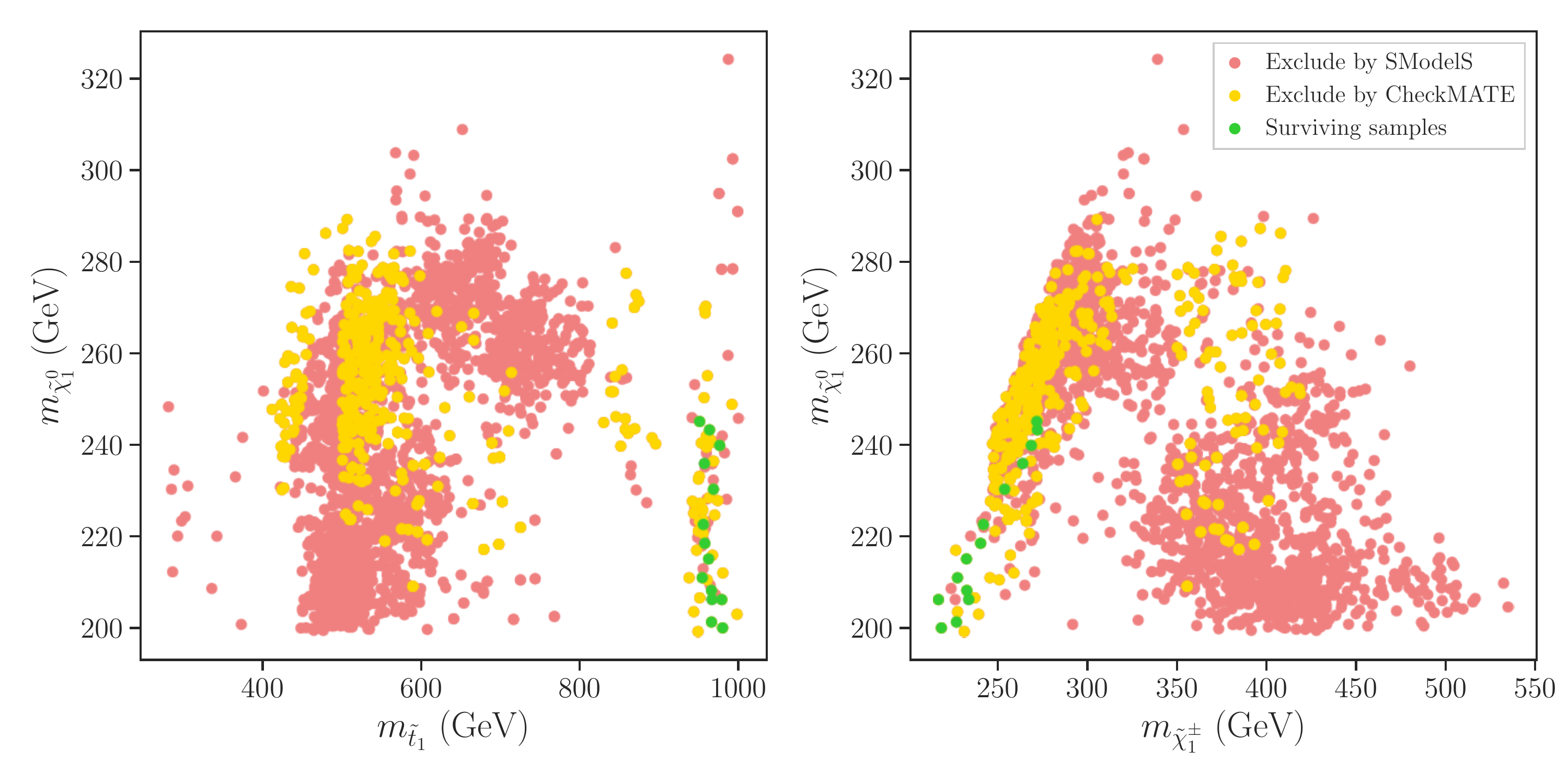}
 \caption{  The samples showing the exclusion of LHC direct searches. All these samples satisfy the constraints from the new CDF II measurement of $M_W$, the combined FNAL and BNL muon $g-2$ results, the B-physics measurements, the SM-like Higgs mass requirement, the upper bound on DM relic density at 95\% CL and the DM direct detection 90\% CL limits. (Color online)
 } 
 \label{fig:lhc}
\end{figure*}

\begin{table}[th] 
\centering
\begin{tabular}{ccccccc}
\toprule
Observable		 & $\mu$		& $\sigma$(exp)	& $\tau$(th)	& Reference \\
\hline
$M_W$ (\gev)  & 80.4335 & 0.0094 & 0.010&  \cite{CDF:2022hxs} \\
$\sin^2\hat{\theta}(M_Z)$ & 0.23121	& 0.00004	& 0.00010	& \cite{ParticleDataGroup:2020ssz} \\
$\Omega_{\chi}h^2$				                 & 0.120		& 0.001		& 0.012		& \cite{Planck:2018vyg} \\
$\Delta a_{\mu}\times 10^{10}$ 			     & 25.1		    & 5.9		    & 2		&\cite{Muong-2:2021ojo}   \\
$B(B\rightarrow X_s\gamma)\times10^4$		     & 3.32		& 0.15		& 0.24		&   \cite{HFLAV:2019otj}  \\
$B(B_s^0 \rightarrow \mu^+\mu^-)\times10^9$	 & 2.69		& 0.36		& 0.29		&  \cite{ParticleDataGroup:2020ssz} \\
$B(B^+\rightarrow \tau^+ \nu_{\tau})\times10^4$   & 1.06	& 0.19		& 0.23 		& \cite{HFLAV:2019otj}  \\
\toprule
\end{tabular}
\caption{ The Gaussian distributed experimental constraints used in the likelihood of the scans. The DM relic density restriction is implemented as an upper limit. \label{tab:experiment}}
\end{table}

\begin{table}[th]
\centering
\begin{tabular}{ccccc}
\toprule
~~~~Parameter~~~~           & ~~~~Minimum~~~~  & ~~~~Maximum~~~~     & ~~~~Prior~~~~ \\
\hline
$m_t$                   & 171.86   & 173.66	&  gaussian \\
$\tan\beta$             & 5        & 60     &  flat \\
$\mu$                   & -2000    & -2000  &  flat/log \\
$M_1$                   & 100      & 2000   &  flat/log \\
$M_2$                   & 100      & 2000   &  flat/log \\
$M_3$                   & 800      & 3000   &  flat/log \\
$M_A$                   & 90       & 2000   &  flat/log \\
$M_{\Tilde{E}_{1,2,3}}$ & 100      & 2000   &  flat/log \\
$M_{\Tilde{L}_{1,2,3}}$ & 100      & 2000   &  flat/log \\
$M_{\Tilde{Q}_{1,2}}=M_{\Tilde{U}_{1,2}}=M_{\Tilde{D}_{1,2}}$ & 500       & 2000        &  flat/log \\
$M_{\Tilde{Q}_{3}}$     & 500      & 2000   &  flat/log \\
$M_{\Tilde{U}_{3}}$     & 500      & 2000   &  flat/log \\
$M_{\Tilde{D}_{3}}$     & 500      & 2000   &  flat/log \\
$A_e=A_{\mu}=A_{\tau}$  & -5000    & 5000   &  flat \\
$A_u=A_d=A_c=A_s$       & -5000    & 5000   &  flat \\
$A_b$                   & -5000    & 5000   &  flat \\
$A_t$                   & -5000    & 5000   &  flat \\
\toprule
\end{tabular}
\caption{ Ranges and priors for input parameters of MSSM adopted in the scans. The SUSY parameters are defined at the scale of $M_{SUSY}=\sqrt{m_{\Tilde{t}_1}m_{\Tilde{t}_2}}$.
All parameters with mass dimension are given in GeV.\label{tab:ranges}}
\end{table}

We adopt the Markov Chain Monte Carlo (MCMC) method based on the Metropolis-Hastings 
algorithm and MultiNest~\cite{Feroz:2008xx} algorithm to perform the scan of MSSM parameter space, using \texttt{EasyScan\_HEP} \cite{Han:2016gvr} . The Gaussian distributed constraints are implemented in the likelihood to guide the scan, which are listed in Table~\ref{tab:experiment} and take the form of 
\begin{equation}
    \ln\mathcal{L}=- \mathop{\sum}_{i} \frac{(\hat{\mu}_i-\mu_i)^2}{2(\sigma_i^2+\tau_i^2)} -\frac{1}{2}\chi^2_{\texttt{HiggsSignals}},
\end{equation}
where $\hat{\mu}_i$ denotes the predicted value of observable $i$, $\mu_i$ is the corresponding central value of experimental measurement, $\sigma_i$ and $\tau_i$ are experimental and theory uncertainties respectively. Besides, the $\chi^2$ value of Higgs measurements derived from \texttt{HiggsSignals} with 3~\gev theoretical uncertainty on SM-like Higgs mass is also included. The ranges and priors of input parameters are shown in Table~\ref{tab:ranges}. We neglect possible flavor violation in the SUSY sector. 

Here we do not intend to do a global fit study (see, e.g., \cite{GAMBIT:2018gjo,Li:2020glc}) with such high dimension parameter space. So we collect all the samples, not only the reserved samples of the algorithms, and then discard the samples excluded by each of above experimental constraints at 95\% confidence level~(CL). Note that the DM relic density restriction is implemented as an upper limit, so as to permit additional non-neutralino component. Moreover, the 90\% CL bounds from the XENON-1T experiment~\cite{XENON:2018voc,XENON:2019rxp} on the spin-independent and spin-dependent DM-nucleon scattering cross-sections are applied, assuming the additional non-neutralino DM component has no interaction with nucleon. Finally, all the samples are further required to pass all the restrictions from \texttt{HiggsBounds}, \texttt{SModelS} and \texttt{CheckMATE}.

\section{Results and discussions} 
\label{sec:result}

We find in our scan that all constraints listed in Table ~\ref{tab:experiment} and $m_{h_1}=125.25\pm0.17~\gev$~\cite{ParticleDataGroup:2020ssz} with 3~\gev theoretical uncertainty can be satisfied simultaneously at $2\sigma$ CL, without obvious conflicts. However, after further applying the 90\% CL spin-independent limits of Xenon1T for DM direct detection, the surviving samples are characterized by light stops, sbottoms and sleptons. Therefore, we
test the samples consisting with the CDF II measurement of $M_W$ at $2\sigma$ CL using \texttt{SModelS}, which implements constraints from SUSY direct searches at LHC by interpreting simplified-model results.
Here the $2\sigma$ CL constraints are applied independently, so the overall confidence level of each displayed point is stricter than $2\sigma$ level~\cite{AbdusSalam:2020rdj}.

Although most of them are excluded by \texttt{SModelS}, there are two kind of samples surviving, as displayed in Fig.~\ref{fig:lhc} by yellow and green bullets. The first case is obviously that the sparticles are relatively heavy. The other case is that the mass spectrum is compressed or in special structure. For example, we find that some surviving samples satisfy the condition of 
\begin{equation}
    m_{\tilde{e}_1,\tilde{\mu}_1} ~>~ m_{\tilde{\chi}_1^\pm}, m_{\tilde{\chi}_2^0} ~>~ m_{\tilde{\tau}_1}. 
    \label{eq:sms}
\end{equation}
Therefore, wino-dominated $\tilde{\chi}_1^\pm$ and $\tilde{\chi}_2^0$ mainly decay to  $\tilde{\tau}_1 \nu_\tau$ and $\tilde{\tau}_1 \tau$ respectively, and then $\tilde{\tau}_1$ decay to $\tau \tilde{\chi}_1^0$.  Meanwhile, sbottom and stop tend to decay to wino instead of bino. As a result, all the sparticle productions at LHC end up in tau-rich final status, which is not commonly used in simplified-models for LHC SUSY searches. 

However, with detailed simulations using \texttt{MG5@NLO}~\cite{Alwall:2014hca} and \texttt{PYTHIA}~\cite{Sjostrand:2006za,Sjostrand:2007gs}, we find that some surviving samples are further excluded by \texttt{CheckMATE}. A robust constraint is from search for dark matter produced in association with bottom or top quarks using  36.1 fb$^{-1}$  data at 13~\tev LHC \cite{ATLAS:2017hoo}. It can exclude stop and sbottom up to 900~\gev, regardless of the structure in Eq.~\ref{eq:sms}. Finally, only the green dots in Fig.~\ref{fig:lhc} can survive. Note that the stop can not be further heavy because the contributions purely from eletroweak sector to $M_W^{\rm SUSY}$ can not reach $2\sigma$ regions of the CDF II measurements, as studied in \cite{Bagnaschi:2022qhb}. Roughly speaking, with $m_{\tilde{t}_1} > 950~\gev$, the wino-dominated chargino has to be lighter than about 350~\gev, and the structure in Eq.~\ref{eq:sms} no longer exists. 
Thus, the mass splitting between $\tilde{\chi}_1^\pm$ and $\tilde{\chi}_1^0$ has to be smaller than about 30~\gev to avoid multilepton searches at the LHC.

\begin{table}[th]
\scriptsize
\centering
\begin{tabular}{ccccccccccccccc}
\toprule
 & $m_t$ & $\tan\beta$ & $\mu$ & $M_1$  &  $M_2$ &  $M_3$ & $M_A$ & $M_{\Tilde{E}_{1,2,3}}$  & $M_{\Tilde{L}_{1,2,3}}$ \\
\hline
\texttt{P1} & 173.642 & 19.7 & 800.6 & 201.1 & 222.0 & 2204.0 & 1743.5 & 601.7 &  210.2 \\ 
\texttt{P2} & 173.635 & 22.7 & 684.9 & 243.6 & 274.9 & 2105.5 & 2894.4 & 531.5 &  210.2 \\ 
\hline
   & $M_{\Tilde{Q}_{1,2}}$ & $M_{\Tilde{Q}_{3}}$     & $M_{\Tilde{U}_{3}}$     & $M_{\Tilde{D}_{3}}$     & $A_{L_{1,2,3}}$  & $A_{Q_{1,2}}$ & $A_b$                   & $A_t$  \\
\hline
\texttt{P1}  &  4023.7 & 963.6 & 2652.5 & 4015.1 & 3408.8 & 3705.7 & 3158.7 & 5513.3\\
\texttt{P2}  & 4414.0 & 951.8 & 2633.0 & 3991.4 & 2573.9 & 2626.5 & 3174.8 & 5277.5\\
\end{tabular}
\begin{tabular}{ccccccccccccccc}
\toprule
 & $M_W$ & $\sin^2\hat{\theta}(M_Z)$  & $\Delta a_{\mu}$ & $B\rightarrow X_s\gamma$ & $B_s^0 \rightarrow \mu^+\mu^-$ & $B^+\rightarrow \tau^+ \nu_{\tau}$\\
\hline
\texttt{P1}  & 80.4064 & 0.2312355 &  2.15$\times10^{-9}$ & 3.580$\times10^{-4}$ & 2.853$\times10^{-9}$ & $7.648\times10^{-5}$\\
\texttt{P2}  & 80.3981 & 0.2312846 & 2.22$\times10^{-9}$ & 3.501$\times10^{-4}$ & 2.942$\times10^{-9}$ & 7.672$\times10^{-5}$\\
\end{tabular}
\begin{tabular}{ccccccccccccccc}
\toprule
 & $m_{h_1}$     & $m_{h_2}$   & $m_{H^{\pm}}$ & $m_{\tilde{\chi}_1^0}$     & $m_{\tilde{\chi}_2^0}$  & $m_{\tilde{\chi}_3^0}$ & $m_{\tilde{\chi}_4^0}$  & $m_{\tilde{\chi}_1^\pm}$ & $m_{\tilde{\chi}_2^\pm}$ & $m_{\tilde{g}}$\\
\hline
\texttt{P1} & 119.7 & 1743.0 & 1745.9 & 200.0 & 218.9 & 804.2 & 808.4 & 218.7 & 809.5 & 2204.0\\
\texttt{P2} & 122.3 & 2894.0 & 2895.9 & 242.0 & 269.8 & 688.9 & 695.6 & 269.5 & 696.4 & 2105.5\\
\toprule
 & $m_{\tilde{t}_1}$      & $m_{\tilde{t}_2}$    & $m_{\tilde{b}_1}$      & $m_{\tilde{b}_2}$  & $m_{\tilde{\tau}_1}$      & $m_{\tilde{\tau}_2}$  & $m_{\tilde{e}_L}$      & $m_{\tilde{e}_R}$  & $m_{\tilde{u}_L}$  & $m_{\tilde{d}_L}$\\
\hline
\texttt{P1} & 980.2 & 2622.9 & 1034.4 & 4038.6 & 212.4 & 604.4 & 215.6 & 603.3 & 4023.4 & 4024.2\\
\texttt{P2} & 978.3 & 2610.9 & 1028.2 & 4014.8 & 261.6 & 535.3 & 265.9 & 533.2 & 4413.7 & 4414.4\\
\end{tabular}
\begin{tabular}{ccccccccccccccc}
\toprule
& $\tilde{g} \to \tilde{b}_1 b$ & $\tilde{g} \to \tilde{t}_1 t$ &
$\tilde{t}_1 \to  \tilde{\chi}_1^\pm b $ & $\tilde{t}_1 \to  \tilde{\chi}_2^0 t $ &  $\tilde{b}_1 \to \tilde{\chi}_1^\pm t $ & $\tilde{b}_1 \to \tilde{\chi}_2^0 b $ & $\tilde{\chi}_2^0 \to \tilde{\nu_{\ell}} \nu_{\ell}$ & ~~\\
\hline
\texttt{P1} & 47.4\% & 52.6\% & 64.6\%  & 30.4\% & 51.9\% & 26.9\% & 92.7\% \\
\texttt{P2} & 47.4\% & 52.6\% & 45.0\%  & 20.8\% & 38.7\% & 20.4\% & 86.2\% \\
\end{tabular}
\begin{tabular}{ccccccccccccccc}
\toprule
& $\tilde{\chi}_1^{\pm} \to \tilde{\nu} \ell^{\pm}$ 
& $\tilde{\nu}\to \tilde{\chi}_1^0 \nu$ 
& $\tilde{\ell}_1 \to  \tilde{\chi}_1^0 \ell $ 
& $\tilde{\chi}_2^{\pm} \to \tilde{\chi}_2^{0} W^{\pm}$
& $\tilde{\chi}_2^{\pm} \to \tilde{\chi}_1^{\pm} h_1$
& $\tilde{\chi}_3^{0} \to \tilde{\chi}_1^{\pm} W^{\mp}$
& ~~
\\
\hline
\texttt{P1} & 94.3\% & 100\% & 100\%  & 30.7\% & 26.5\% & 57.2\%\\
\texttt{P1} & 89.6\% & 100\% & 100\%  & 30.5\% & 25.0\% & 56.9\%\\
\toprule
\end{tabular}
\caption{ Two benchmark points satisfying all relevant collider and DM constraints. All masses and trilinear couplings are in unit of GeV while the decays refer to the branching ratios. For point \texttt{P1}  both the W-boson mass and muon $g-2$ are in the $2\sigma$ ranges of the experimental values. \label{tab:bks}}
\end{table}
In Tab ~\ref{tab:bks}, we list two benchmark points that satisfy relevant constraints including LHC direct searches at $2\sigma$ CL. Point \texttt{P1} is consistent with the new $M_W$ measurements in $2\sigma$ CL, but has small SM-like Higgs mass. Considering conservative requirement on SM-like Higgs mass, we give a second benchmark point \texttt{P2}, which has $m_h=122.3~\gev$ but $M_W^{\rm SUSY}$ lower than $2\sigma$ bound of the new $M_W$ measurements. 
From the view of LHC SUSY searches, the two point have similar mass spectrums.
Thus, the R-value, ratio of 95\% lower limit on the number of LHC signal events to corresponding experimentally measured 95\% CL limit, obtained from \texttt{CheckMATE} is 0.84 for point \texttt{P1}  and 0.85 for point \texttt{P2}. 
The dominated contributions are $pp\to\tilde{t}_1\tilde{t}_1$ and $pp\to\tilde{b}_1\tilde{b}_1$ processes, with similar final status.
Therefore, they are on the edge of exclusion. With a more substantial scan, the R-value could be slightly reduced, but there is no doubt that the region around this benchmark point will be fully covered with higher luminosity at LHC.

\begin{figure}[!t]
\centering
 \includegraphics[width=0.5\textwidth]{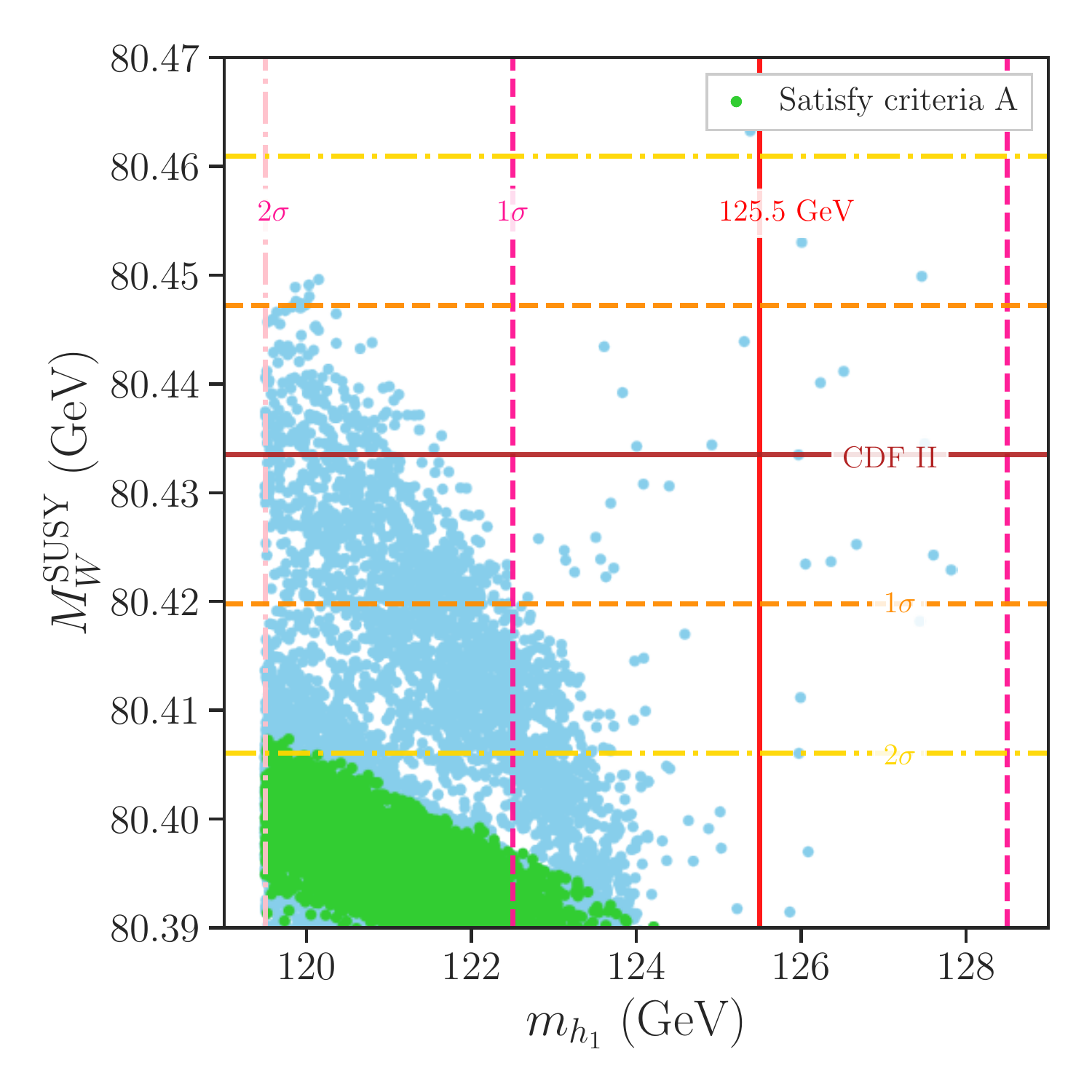}
 \caption{ 
 The samples satisfying constraints from the combined FNAL and BNL muon $g-2$ results, the B-physics measurements, the SM-like Higgs mass measurement and the DM detections, with green bullets satisfying  $m_{\tilde{t}_1}>0.95~\tev$, $m_{\tilde{b}_1}>1~\tev$,
 $m_{\tilde{g}}>1.5~\tev$, $m_{\tilde{q}}>2~\tev$ and $m_{\tilde{\chi}_1^\pm} - m_{\tilde{\chi}_1^0} < 30~\gev$, namely the current LHC direct search constraints. We adopt 0.010~\gev and 3~\gev theoretical uncertainties for W-boson mass and SM-like Higss mass calculations. (Color online)
 }
 \label{fig:mh}
\end{figure}

\begin{figure*}[!th]
 \centering
 \includegraphics[width=0.98\textwidth]{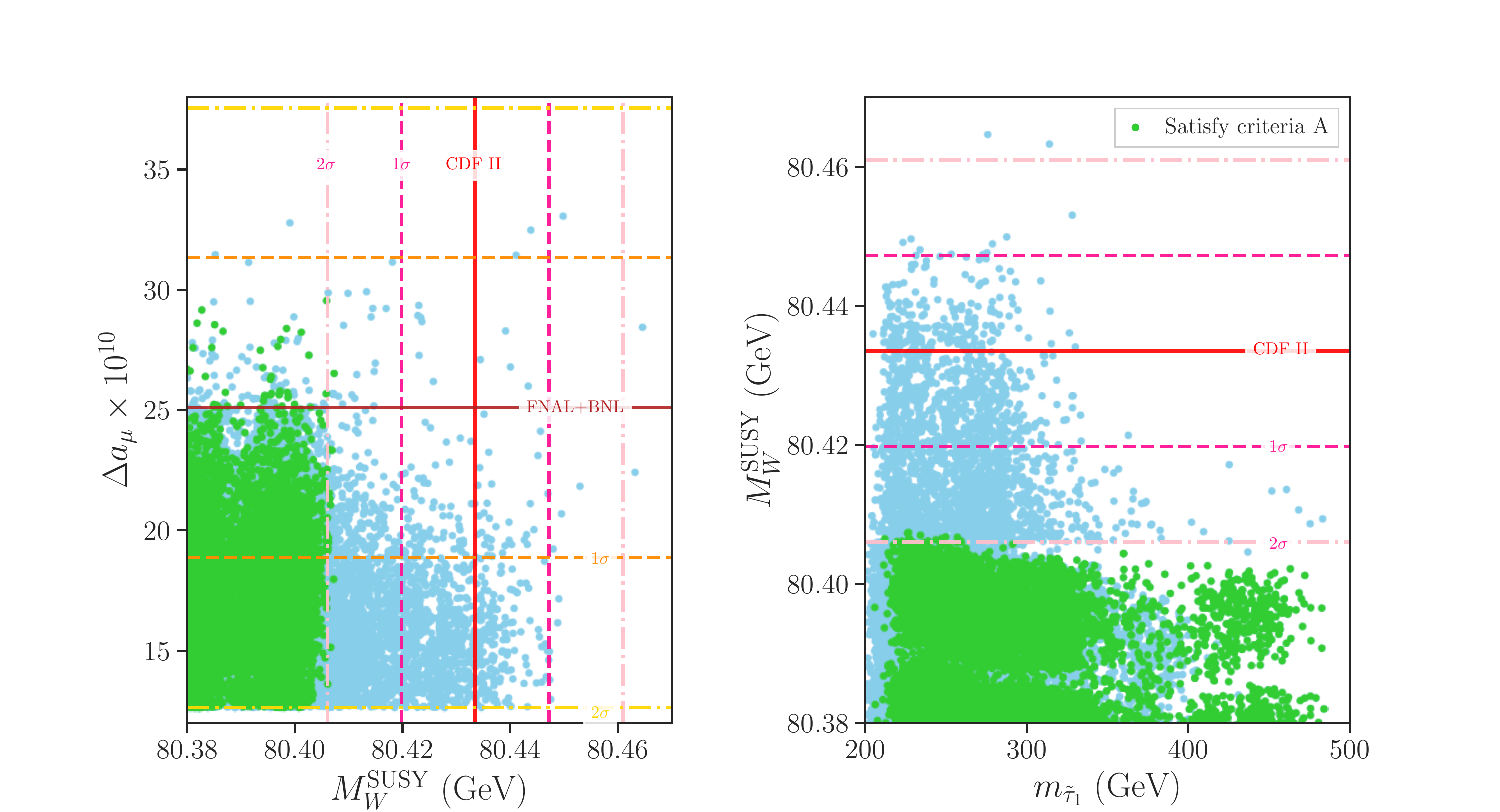}
 \caption{  The samples satisfying constraints from the combined FNAL and BNL muon $g-2$ results, the B-physics measurements, the SM-like Higgs mass measurement,  the upper bound on DM relic density at 95\% CL and the DM direct detection 90\% CL limits. The green bullets further satisfy the current LHC direct search constraints. The horizontal and vertical solid, dashed and dot-dashed lines indicate the current experimental central values, $1\sigma$ and $2\sigma$ ranges of the corresponding observables. (Color online)}
 \label{fig:gm2}
\end{figure*}

From the benchmark points, we see that our surving samples are also on the edge of $2\sigma$ limits from the LHC SM Higgs measurements and the CDF II $M_W$ measurements. Requiring $m_{\tilde{t}}>1~\tev$, it is hard to push $M_W^{\rm SUSY}$ over 80.4~\gev, which is consistent with a precious study~\cite{Heinemeyer:2013dia}, and to obtain a correct SM-like Higgs mass in the MSSM. In Fig.~\ref{fig:mh}, we display the tension between $M_W^{\rm SUSY}$ and $m_{h_1}$. We define a criteria 
\begin{align*}
    \texttt{Criteria A}:~~& m_{\tilde{t}_1}>0.95~\tev,~~ m_{\tilde{b}_1}>1~\tev, \\
    & m_{\tilde{g}}>1.5~\tev,~~ m_{\tilde{q}}>2~\tev, \\ 
    & m_{\tilde{\chi}_1^\pm} - m_{\tilde{\chi}_1^0} < 30~\gev
\end{align*}
according to results of \texttt{CheckMATE}. The samples satisfying this criteria are safe from the LHC direct searches. We can see from Fig.~\ref{fig:mh} that the LHC direct searches, the SM-like Higgs measurement and the CDF II $M_W$ measurement can be simultaneously satisfied at $2\sigma$ CL, but near the edge of $2\sigma$ range, with loose theoretical uncertainties. Besides, heavy stop in the MSSM may also cause vacuum stability problem, which could be avoid in extensions of the Higgs sector, such as the NMSSM.

\begin{figure*}[th]
 \centering
 \includegraphics[width=0.9\textwidth]{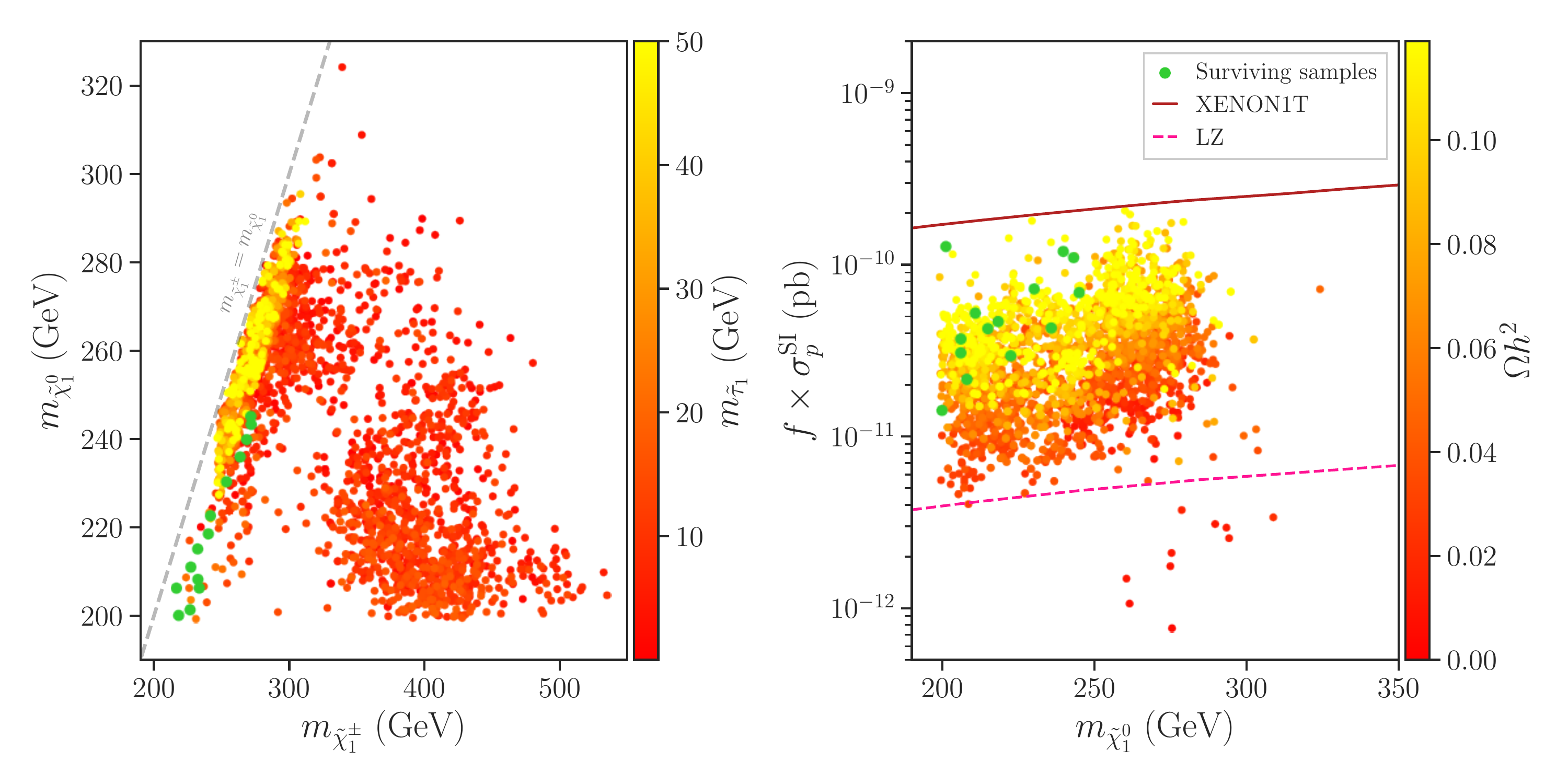}
 \caption{  Same as Fig.~\ref{fig:lhc}, but showing DM properties. The green samples satisfy all constraints including the LHC SUSY direct searches.
 The upper bound of color map in the left panel is manually set, and $m_{\tilde{\tau}_1}-m_{\tilde{\chi}_1^0} $ can reach to 200~\gev. In the right panel, the 90\% CL exclusion limits from the XENON1T and the projected reach of LZ are also presented, and $f$ indicates the ratio of the predicted relic density to the value observed by Planck. (Color online) }
 \label{fig:before_lhc}
\end{figure*}

For the surviving samples with $M_W^{\rm SUSY}>80.406$~\gev, the stop and sbottom contributions described in Eq.~\ref{eq:stop_contr} dominate the SUSY correction to $M_W^{\rm SUSY}$, while the contributions of stau constitutes about 10\% of that. Nevertheless, a light stau is still needed, because its small contribution helps to push a sample into the $2\sigma$ range of $M_W$. In addition, the universal slepton mass is required to be light to achieve agreement with the FNAL measurement of muon $g-2$. These features can be seen in Fig.~\ref{fig:gm2}. 

In the left panel of Fig.~\ref{fig:gm2}, we plot the surviving samples in the plane of $M_W^{\rm SUSY}$ versus $\Delta a_{\mu}$.  The horizontal and vertical solid, dashed and dot-dashed lines indicate the current central value, $1\sigma$ limits and $2\sigma$ limits of the corresponding observable.
The samples satisfying \texttt{Criteria A}, i.e. the LHC SUSY search limits, are colored by green.
We can see that the few surviving samples in the $2\sigma$ region of the new CDF II $M_W$ measurement can also agree very well with the FNAL muon $g-2$ measurement, due to the fact that they both favour light sleptons. Actually, the correlations between $M_W^{\rm SUSY}$ and other observables listed in Tab.~\ref{tab:experiment} are rather weak, except for $\sin^2\hat{\theta}(M_Z)$ in which SUSY corrections are also proportion to $\Delta\rho$. We also check that requiring lightest neutralino to be the only DM candidate, i.e. applying also lower bound on DM relic density, the results will barely change.

The right panel of Fig.~\ref{fig:gm2} shows the relation between the SUSY prediction of $W$-boson mass and the stau mass. Note that the point density in the plots does not have statistical meaning, because we have several scans focusing on special regions. For instance, the compressed region of ($m_{\tilde{\tau}_1}, m_{\tilde{\chi}_1^{\pm}}, m_{\tilde{\chi}_1^{0}}$) is heavily sampled, because it is promising to avoid constraints of LHC direct searches. We can see that before implementing constraints of LHC SUSY direct searches, a large $M_W^{\rm SUSY}$ can be achieved with a light stau. However, with requiring \texttt{Criteria A}, the maximal value of $M_W^{\rm SUSY}$ increases rather mildly with $m_{\tilde{\tau}_1}$ decreasing. As a result, in the $2\sigma$ region of $M_W$, we see that $m_{\tilde{\tau}_1} <  260~\gev$.

In such a mass region, DM can achieve desired relic abundance through chargino or/and stau co-annihilation mechanisms. Therefore, the lightest stau $\tilde{\tau}_1$ or/and chargino $\tilde{\chi}_1^{\pm}$ have small mass splitting to the DM, where $\tilde{\chi}_1^{\pm}$ is always wino-dominated in our results, as shown in the left panel of Fig.~\ref{fig:before_lhc}.
The final surviving samples, colored by green, have degenerated $\tilde{\tau}_1$, $\tilde{\chi}_1^{\pm}$ and $\tilde{\chi}_1^{0}$.
In the right panel of Fig.~\ref{fig:before_lhc}, we show the scaled spin-independent neutralino-proton cross-section of samples further satisfying the CDF II measurements at $2\sigma$ CL.
We can see that the surviving samples can be fully covered by the projected LUX-ZEPLIN~(LZ) experiment~\cite{LUX-ZEPLIN:2018poe}. 
 Even without the LHC SUSY constraints, the LZ experiment could exclude most of the samples, except for several samples that give tiny DM relic density via extremely degenerating mass spectrum.

\section{Conclusions}
\label{sec:conclusion}
The precision measurements of electroweak observables and muon $g-2$ test the SM in depth and allow for an indirect probe of new physics beyond the SM. 
The new CDF II measurement of $W$-boson mass, which showed a 7$\sigma$ deviation from the SM prediction, and the FNAL measurement of the muon $g-2$, which showed a 4.2$\sigma$ deviation (combined with the BNL result) from the SM,  strongly indicate new physics.   
In this work we scrutinized the implication of both measurements on low energy supersymmetry. We explored the parameter space of the MSSM considering 
various experimental constraints from colliders and DM detections.
We found that in the allowed parameter space, the MSSM contributions can simultaneously satisfy both measurements at $2\sigma$ level, but quite near the edge.  
With a more comprehensive scan or a global fit, it is possible to improve agreement with the measurements.
The favored parameter space, characterized by 
a compressed spectrum between bino, wino and stau, with the stop being around 1 TeV, may be covered in the near future LHC searches.  

\acknowledgments

We thank Lei Wu for helpful discussions. 
This work was supported by the National Natural Science Foundation of China 
(NNSFC) under grant Nos. 11821505, 12075300  and 12105248,  
by the Key Research Project of Henan Education Department for colleges and universities under grant number 21A140025,
by Peng-Huan-Wu Theoretical Physics Innovation Center (12047503),
by the CAS Center for Excellence in Particle Physics (CCEPP), 
by the CAS Key Research Program of Frontier Sciences, 
 and by a Key R\&D Program of Ministry
of Science and Technology of China under number 2017YFA0402204, and by the Key Research Program of the Chinese Academy of Sciences, Grant NO. XDPB15.

\bibliographystyle{unsrt}
\bibliography{refs}

\end{document}